\documentclass[pra,aps,twocolumn,epsf,amsfonts]{revtex4}

\usepackage{graphicx}
\usepackage{epsfig}
\def\qed{\leavevmode\unskip\penalty9999 \hbox{}\nobreak\hfill
     \quad\hbox{\leavevmode  \hbox to.77778em{%
               \hfil\vrule   \vbox to.675em%
               {\hrule width.6em\vfil\hrule}\vrule\hfil}}
     \par\vskip3pt}

\def\ibb #1{\leavevmode\hbox{\kern.3em\vrule
     height 1.5ex depth -.1ex width .4pt\kern-.3em\rm#1}}

\newtheorem{lemma}{Proposition}

\def\ket #1{\vert #1\rangle}

\def\ketbra #1#2{\vert #1\rangle \! \langle #2\vert}

\begin{document}

\title{On the irreversibility of entanglement distillation}
\author{Karl Gerd H. Vollbrecht$^1$, Reinhard F. Werner$^1$, and Michael M. Wolf$^{1,2}$}
\address{$^1$ Institut f\"ur Mathematische Physik, Mendelssohnstr.3,
D-38106 Braunschweig, Germany\\ $^2$ Max-Planck-Institut f\"ur
Quantenoptik, Hans-Kopfermann-Str. 1, D-85748 Garching, Germany}
\date{\today}

\begin{abstract}
We investigate the irreversibility of entanglement distillation
for a symmetric ($d-1$) parameter family of mixed bipartite
quantum states acting on Hilbert spaces of arbitrary dimension
$d\times d$. We prove that in this family the entanglement cost is
generically strictly larger than the distillable entanglement,
such that the set of states for which the distillation process is
asymptotically reversible is of measure zero. This remains true
even if the distillation process is catalytically assisted by pure
state entanglement and  every operation is allowed, which
preserves the positivity of the partial transpose. It is shown,
that reversibility occurs only in cases where the state is
quasi-pure in the sense that all its pure state entanglement can
be revealed by a simple operation on a single copy. The reversible
cases are shown to be completely characterized by minimal
uncertainty vectors for entropic uncertainty relations.
\end{abstract}

\pacs{03.67.-a, 03.65.Ca, 03.65.Ud}

\maketitle

\section{Introduction}
One of the paradigmatic innovations of quantum information theory
is to consider entanglement not only as an apparently paradoxical
feature of distributed quantum systems, but rather as a physical
resource for quantum information processing tasks. This resource
point of view naturally raises the question, how to quantify the
entanglement.

For pure bipartite quantum states there exists, under reasonable
assumptions, a unique measure of entanglement (cf. \cite{dhr02}),
which can in addition easily be calculated. For mixed states,
however, there is a large variety of inequivalent entanglement
measures, which are either extremely hard to calculate, or do not
satisfy some of the ``reasonable assumptions''. Two of these
measures, however, stand out due to there physical meaning: the
{\it entanglement cost} $E_c$, and the {\it distillable
entanglement} $E_D$. $E_c$ is the amount of pure state
entanglement asymptotically needed in order to prepare a given
mixed state, whereas the  distillable entanglement is the amount
of pure state entanglement that can asymptotically be extracted
from it (cf. \cite{spaeter}). Both, preparation and distillation
process, are assumed to use only local operations and classical
communication (LOCC).

The present paper is devoted to the question to what extent
entanglement distillation is an (ir-)reversible operation. We will
give a complete characterization of the reversibility properties
of the entire set of non-trivial distillable mixed states for
which the (entanglement assisted) distillable entanglement is
known so far.

 It was recently proven that there are indeed examples, for which
$E_D$ is strictly smaller than $E_c$, hence entanglement
distillation is not reversible in general \cite{Cirac, VC01,VC02,
HSS0}. However, it is yet unclear, whether this is an exceptional
feature of specific quantum states, or a generic property of all
mixed states. The particular variational and asymptotic nature of
$E_c$ and $E_D$ makes this  an intrinsically hard problem, even if
we have not to calculate both measures explicitly. Although a
general answer to this question seems  to be out of reach at the
moment, we are able to completely classify a $(d-1)$ parameter
family of non-trivial mixed states in arbitrary Hilbert space
dimensions $d\times d$ with respect to their reversibility
properties.

For the considered class of states, irreversibility turns out to
be generic. This remains true even if the distillation process is
catalytically assisted by pure state entanglement and the class of
allowed distillation operations is considerably enlarged.
Moreover, all the reversible cases correspond to pure states which
are hidden in a mixture, that can, however, be broken up by means
of local operations. Surprisingly, the equality $E_c\geq E_D$ is
for the states under consideration essentially equivalent to a
discrete entropic uncertainty relation, such that the cases of
reversibility are completely characterized by vectors of minimal
uncertainty. The Appendix shows how to explicitly calculate the
amount of {\it undistillable entanglement} $(E_c-E_D)$ for some
examples.

Before specifying and proving these results, we will begin with
providing some prerequisites and recalling recent results.

\section{Preliminaries}
\subsection{States under consideration}\label{states}

Let $\{\Psi_{kl}\}$ with $k,l = 0,\ldots ,d-1$ be an orthonormal
basis of maximally entangled states in ${\cal
H}_{AB}={\mathbb{C}}^d\otimes{\mathbb{C}}^d$,  given by
\begin{equation}\label{UPsi}
|\Psi_{kl}\rangle = \frac1{\sqrt{d}} \sum_{j=0}^{d-1}
|j\rangle\otimes U_{kl}|j\rangle,\quad U_{kl}= \sum_{r=0}^{d-1}
\eta^{rl} |k + r\rangle\langle r|,
\end{equation}where addition inside the ket is modulo $d$ and $\eta=e^{\frac{2\pi
i}d}$.

We will in the following consider states of the form
\begin{equation}\label{rholambda}
\rho_{\lambda} = \sum_{l=0}^{d-1}
\lambda_l\;|\Psi_l\rangle\langle\Psi_l|,\quad \Psi_l=\Psi_{0l},
\end{equation} which form a $(d-1)$ parameter family of rank deficient states,
characterized by the probability vector
$\lambda\in{\mathbb{R}}^d$. Note that we could equivalently choose
$\Psi_l= \Psi_{g + f l,l}$ with arbitrary integers $g,f$, or any
other set of states, which differs from $\{\Psi_l\}$ only by the
choice of local bases. It is readily verified that
$\rho_{\lambda}$ is entangled unless $\lambda_l\equiv\frac1d$
(cf.\cite{VWhash}). Like all states, which are diagonal with
respect to the basis $\{\Psi_{kl}\}$, every state $\rho_{\lambda}$
commutes with any element of the abelian symmetry group
$G=\{U_{k,l}\otimes U_{k,-l}\}$ \cite{VWhash}. In particular,
$\rho_{\lambda}$ is invariant under the LOCC twirl operation
${\cal T}(\rho)=d^{-2}\sum_{g\in G} g^*\rho g$, which maps any
state $\rho$ onto a symmetric state ${\cal T}(\rho)$, which is
diagonal with respect to $\{\Psi_{kl}\}$. We will in the following
make extensive use of the symmetry properties of $\rho_{\lambda}$
and the particular form of $\Psi_l$.


\subsection{Entanglement measures and state transformations}
For pure bipartite states there exists, under reasonable
assumptions, a unique measure of entanglement: the von Neumann
entropy of the reduced state $E(\varphi)=S\big({\rm
tr}_A({|\varphi\rangle\langle\varphi|})\big)$, where
$S(\rho)=-\sum_i p_i\log p_i$ is the Shannon entropy of the
spectrum $\{p_i\}$ of $\rho$. For simplicity we will use the same
notation for the Shannon and von Neumann entropy, i.e.,
$S(\rho)=S(\{p_i\})$.

Every pure state $\varphi$ can by means of LOCC operations be
transformed into maximally entangled states $\psi$ in an
asymptotically reversible manner, such that the number of
maximally entangled states per copy of $\varphi$ is given by
$E(\varphi)$. That is, we can locally transform $\psi^{\otimes n
E(\varphi)} \Rightarrow \varphi^{\otimes n} \Rightarrow
\psi^{\otimes n E(\varphi)}$ for $n\rightarrow\infty$, where
$\psi$ is any maximally entangled two qubit state.

For a mixed state $\rho$ the amount of pure state entanglement
which is asymptotically required for the preparation of $\rho$ and
the amount that can be extracted from it, need not be the same.
However, it is from a physical point of view obvious, that the
entanglement cost $E_c$ is in general larger than or equal to the
distillable entanglement $E_D$, where $E_D$ and $E_c$ are the
asymptotically optimal rates in
\begin{equation}\label{mixedconversion}
  \psi^{\otimes n E_c} \Rightarrow \rho^{\otimes n} \Rightarrow
\psi^{\otimes n E_D},\quad\text{for}\ n\rightarrow\infty.
\end{equation}In fact, it was proven in \cite{Cirac, VC01,VC02}, that there are
undistillable ($E_D=0$) as well as distillable $(E_D>0)$ cases,
for which $E_c>E_D$, such that there is  some {\it undistillable
entanglement} contained in the state, which makes the distillation
process irreversible. One of these examples investigated in
\cite{Cirac} is in fact $\rho_{\lambda}$ for the case $d=2$.

It is well known \cite{terhal} that
$E_c(\rho)=\lim_{n\rightarrow\infty} E_f(\rho^{\otimes n})/n$ is
the asymptotic {\it entanglement of formation} $E_f(\rho)=\inf
\sum_i q_i E(\varphi_i)$, where the infimum is taken over all pure
state decompositions  $\rho=\sum_i
q_i|\varphi_i\rangle\langle\varphi_i|$. For symmetric states like
$\rho_{\lambda}$ this computation can be simplified \cite{VWsym},
such that $E_f=\text{co}\; \epsilon(\rho)$ is the convex hull of
the function
\begin{equation}\label{epsilon}
\epsilon(\rho)=\inf\big\{E(\varphi)|\rho={\cal
T}(|\varphi\rangle\langle\varphi|)\big\}.
\end{equation}
Moreover, it was shown in \cite{VWhash, Rains2} that the
distillable entanglement
 for states of the form $\rho_{\lambda}$ is bounded from above by
$E_D(\rho_{\lambda})\leq\log d-
S(\{\lambda_i\})=E_D^+(\rho_{\lambda})$ \cite{distent}. Here
$E_D^+$ is the {\it PPT entanglement assisted} distillable
entanglement. That is, the class of allowed operations leading to
the optimal rate $E_D^+$ is enlarged from LOCC to maps preserving
the positivity of the partial transpose (PPT) \cite{PPT}, and the
process is in addition assisted by loaned pure-state entanglement.

\subsection{Discrete uncertainty relations}
 The well known Heisenberg
uncertainty relation states that a function and its Fourier
transform cannot both be highly concentrated. Analogous  relations
can  also be formulated for the case of a finite dimensional
vector $c\in{\mathbb{C}}^d$ ($c\neq 0$) and its discrete Fourier
transform $\hat{c}$, $\hat{c}_k=\frac1{\sqrt{d}}\sum_l \eta^{kl}
c_l$. It is  known, for instance from classical signal recovery
\cite{CRS} that
\begin{equation}\label{DonohoStark}
|\text{supp}(c)|\cdot|\text{supp}(\hat{c})|\geq d,
\end{equation}where $|\text{supp}(c)|$ is the number of non-zero
components of $c$. Equality in Eq.(\ref{DonohoStark}) is attained
iff \cite{CRS}
\begin{equation}\label{DSequality}
c_l\;=\;\alpha\; \eta^{\beta l}\; \delta_{0,(l+\gamma)\text{mod}
d_1},
\end{equation}
where $\beta, \gamma$ are arbitrary integers, $\alpha$ is any
proportionality constant and $d_1$ is a factor of $d$, such that
$d_1 d_2 = d$.

Instead of quantifying the ``concentration'' of $c$ and $\hat{c}$
by their supports, we could as well use the entropy, and it was
proven in \cite{Maavier}, that if $c$ is normalized, then
\begin{equation}\label{Maassen}
S\big(\{|c_l|^2\}\big)+S\big(\{|\hat{c}_k|^2\}\big)\geq \log d.
\end{equation}
The case of equality in Eq.(\ref{Maassen}) was discussed in
\cite{iff}, where it was shown that in this case all non-zero
$c_l$ as well as all $\hat{c}_k\neq 0$ have to have the same
moduli, such that Eq.(\ref{Maassen}) reduces to
Eq.(\ref{DonohoStark}) and the complete set of minimal uncertainty
vectors is again given by Eq.(\ref{DSequality}).

We will discuss in the following section, how the entropic
uncertainty inequality is for the set of states $\rho_{\lambda}$
related to the irreversibility of entanglement distillation.

\section{Generic irreversibility}
Our aim is to give a complete characterization of the cases of
equality in $E_c(\rho_{\lambda})\geq E_D(\rho_{\lambda})$ and
$E_c(\rho_{\lambda})\geq E_D^+(\rho_{\lambda})$. We have therefore
to find a sufficiently close estimate for the entanglement cost,
which is considerably simplified by the symmetry properties of
$\rho_{\lambda}$.

First, we can  utilize a result of  \cite{Cirac} in order to show
that $E_c=E_f$: if there is an isometry $V:{\cal
H}_A\rightarrow{\cal H}_A\otimes{\cal H}_B$, which is such that
${\cal M}(X)= {\rm tr}_B(VXV^*)$ is an {\it entanglement breaking}
map, then $E_c(\rho)=E_f(\rho)$ holds for every $\rho$ with
support in $VV^*$. The general form \cite{shor} of an entanglement
breaking map, which is such that $\text{id}\otimes{\cal M}$
``breaks'' the entanglement of every input state, is given by
\begin{equation}\label{entbreaking}
{\cal M}(X)=\sum_j \sigma_j \;{\rm tr}(F_j X),
\end{equation}
where $\{F_j\}$ is a POVM ($F_j\geq 0,\;\sum_j F_j={\bf 1}$) and
$\{\sigma_j\}$ is a set of density operators.

Let us  choose $V:|l\rangle\mapsto |\Psi_l\rangle$, such that
$\rho_{\lambda}$ is clearly in the support of $VV^*$. Then ${\cal
M}(X)=\sum_l \langle \phi_l|X|\phi_l\rangle |l\rangle\langle l|$,
with $|\phi_l\rangle=\frac1{\sqrt{d}}\sum_j \eta^{-lj} |j\rangle$
has the required property, and hence, the entanglement cost is
indeed equal to the entanglement of formation \cite{HSS}.

Utilizing that $\rho_{\lambda}$ has a  symmetry group $G$ of local
unitaries, we can furthermore simplify the calculation of $E_f$,
such that we have  the chain of (in-)equalities:
\begin{equation}\label{chain}
E_D\leq E_D^+\leq E_c=E_f=\text{co}\; \epsilon\leq \epsilon.
\end{equation}
Now note that $\text{co}\;\epsilon$ is strictly smaller than
$\epsilon$ only on affine pieces. Since
$E_D^+(\rho_{\lambda})=\log d- S(\{\lambda_l\})$ is, however,
strictly convex, i.e., nowhere affine,
$E_D^+(\rho_{\lambda})=\text{co}\;\epsilon(\rho_{\lambda})$
implies that also
$E_D^+(\rho_{\lambda})=\epsilon(\rho_{\lambda})$. We can therefore
disregard taking the convex hull, since it does not alter the set
of states for which equality is obtained.

In order to specify the function $\epsilon(\rho_{\lambda})$ we
need first a characterization of pure states $\varphi$, which are
mapped onto $\rho_{\lambda}$ by the twirl operation ${\cal
T}(|\varphi\rangle\langle\varphi|)$. Since $\cal T$ preserves
every expectation value with respect to any element of
$\{\Psi_{kl}\}$, and the latter form an orthonormal basis in
Hilbert space, we have that $\varphi$ has to be of the form
\begin{equation}\label{varphic}
|\varphi\rangle = \sum_{l=0}^{d-1} c_l
|\Psi_l\rangle,\quad\text{with}\ |c_l|^2=\lambda_l.
\end{equation} Straightforward calculation shows then, that the
entanglement of $\varphi$ is given by
$E(\varphi)=S\big(\{|\hat{c}_k|^2\}\big)$, where $\hat{c}$ is the
discrete Fourier transform of $c$.

Combining all these observations we obtain that cases of equality
in $E_c(\rho_{\lambda})\geq E_D^+(\rho_{\lambda})$ are exactly
given by the minimal uncertainty vectors for the discrete entropic
uncertainty relation, i.e.,
$$
S\big(\{|c_l|^2\}\big)+S\big(\{|\hat{c}_k|^2\}\big)= \log d\
\Rightarrow\ E_c(\rho_{\lambda})= E_D^+(\rho_{\lambda}),\nonumber
$$where $\lambda_l=|c_l|^2$. Due to Eq.(\ref{DSequality}) the
vectors of minimal uncertainty correspond to a discrete set of
states $\rho_{\lambda}$. In other words, the irreversibility of
the process of entanglement distillation is generic even in the
case where the distillation process is catalytically assisted by
 pure entangled states and the allowed operations are PPT preserving rather than LOCC. Moreover, in prime
dimensions Eq.(\ref{DSequality}) gives only rise to trivial
solutions corresponding to pure states, so that in this case every
mixed state $\rho_{\lambda}$ contains some {\it undistillable
entanglement}.

We will in the following section have a closer look at the
reversible cases appearing in non-prime dimensions.

\begin{figure}\epsfig{file=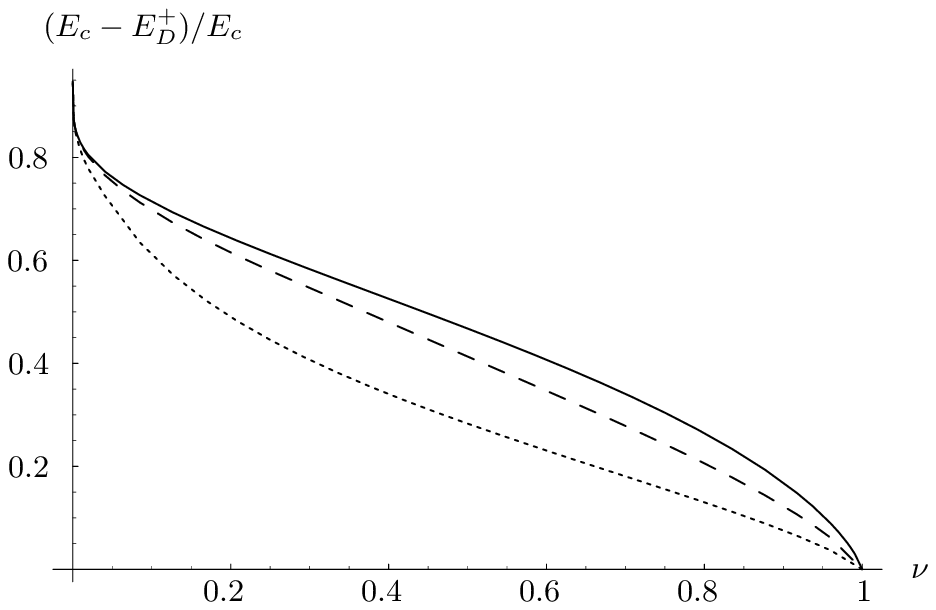,width=8.5cm}
\caption{\label{fig1}The relative amount of {\it undistillable
entanglement} contained in $\rho_{\lambda(\nu)}$ for $d=2,10,100$
(solid, dashed and dotted curve respectively). Whereas near the
separable boundary ($\nu=0$) almost all the entanglement is
undistillable, close to the maximally entangled state ($\nu=1$)
most of the entanglement can be revealed by entanglement
distillation.}
\end{figure}

\section{Quasi-pure states}
It is not difficult to see, that there are entangled mixed states
for which the distillation and preparation processes are
asymptotically reversible. A simple example are pure states
``hidden'' in a mixture, which can, however, be broken up by means
of LOCC operations without destroying entanglement. The simple
idea is to add a Hilbert space ${\cal H}_C$, which contains a
locally measurable {\it tag} $t$ for every entangled state
$\Phi_t\in{\cal H}_A\otimes{\cal H}_B$ appearing in the mixture:
\begin{equation}\label{schummel}
\tau = \sum_t p_t |\Phi_t\rangle\langle\Phi_t|\otimes
|t\rangle\langle t|.
\end{equation}
The {\it locally distinguishable} states $|t\rangle\in{\cal H}_C$
may either correspond to one or to both parties. In the latter
case they have to be products such that they do not contain any
entanglement. It is then easy to see, that
$E_c(\tau)=E_D(\tau)=\sum_t p_t E(\Phi_t)$, since the pure state
$\Phi_t$ can by construction be identified with unit efficiency by
local measurements on ${\cal H}_C$. As states of the form $\tau$,
which have already appeared in \cite{schummel}, are in this sense
essentially (hidden) pure states, we will call them {\it
quasi-pure}.

Let us now turn to the case
$E_c(\rho_{\lambda})=E_D^+(\rho_{\lambda})$ again. For simplicity
we will change the local bases and choose
$\Psi_l=\Psi_{l0}=\frac1{\sqrt{d}}\sum_j |j,j+l\rangle$  now.
Eq.(\ref{DSequality}) tells us that reversibility in dimension
$d=d_1 d_2$ occurs only if the state is up to local unitaries of
the form $\rho_{\lambda}$ with
$\lambda_l=\frac1{d_2}\delta_{0,l\;\text{mod} d_1}$. Writing down
this state leads after some simple substitutions to
\begin{equation}\label{tensorschummel}
\rho_{\lambda}=
|\Psi_{00}^{(d_1)}\rangle\langle\Psi_{00}^{(d_1)}|\otimes
\sum_{k=0}^{d_2-1}\frac1{d_2}|\Psi_{k0}^{(d_2)}\rangle\langle\Psi_{k0}^{(d_2)}|,
\end{equation}where the upper indices label the dimension of the
respective states. The tensor product in Eq.(\ref{tensorschummel})
corresponds to a split of the form ${\cal H}_{A_1B_1}\otimes {\cal
H}_{A_2B_2}$ rather than to an $A|B$ split. As pointed out in
Sec.\ref{states} the second part in Eq.(\ref{tensorschummel}) is,
however, a separable state, such that each reversible
$\rho_{\lambda}$ corresponds to the trivial case of a  quasi-pure
state for which all $\Phi_t$ are the same. It is evident that in
these cases $E_D=E_D^+$, since the states are already distilled
after discarding the separable part.

\section{Conclusion}

We have discussed the irreversibility of entanglement distillation
for the entire set of non-trivial distillable mixed states for
which the (entanglement assisted) distillable entanglement is
known so far. The result is supporting evidence for the conjecture
that irreversibility in the  distillation process is a generic
property of mixed states. As a byproduct we rederived the discrete
entropic uncertainty relation (for complementary observables) from
the simple statement that entanglement is non-increasing under
LOCC operations. Within the considered set of states reversible
cases are thus characterized by minimal uncertainty vectors, which
in turn were shown to correspond to {\it quasi-pure states}.

\section*{Acknowledgement} MMW is grateful to H. Maassen for valuable comments
on entropic uncertainty relations.
 Funding
by the European Union project EQUIP (contract IST-1999-11053) and
financial support from the DFG (Bonn) is gratefully acknowledged.

\appendix
\section{} Here, we will show how to explicitly calculate the
amount of {\it undistillable entanglement} for some states
$\rho_{\lambda}$. We can in particular circumvent the calculation
of the infimum in Eq.(\ref{epsilon}), if all the eigenvalues
$\lambda_l$ except one are equal:
\begin{lemma}
The entanglement of formation of the state $\rho_{\lambda(\nu)}$
with
$\lambda(\nu)=\{\nu+\frac{1-\nu}{d},\frac{1-\nu}{d},\frac{1-\nu}{d},\dots\}$
is equal to the entanglement of formation of an isotropic state
$\sigma(f)=f |\Psi_0\rangle\langle\Psi_0| +
\frac{1-f}{d^2-1}({\bf{1}}-|\Psi_0\rangle\langle\Psi_0|)$ with
$f=\nu+\frac{1-\nu}{d}$. The entanglement of the latter was
calculated in \cite{TV}.
\end{lemma}
This result is implied by the general extension method for
isotropic states discussed in \cite{VWsym}. We will, however, give
here a simpler proof, that is  based on the fact that for states
diagonal w.r.t. $\{\Psi_{kl}\}$ as well as for isotropic states
there exist a twirl operation $\cal T$ resp. ${\cal{T}}^{U\otimes
{\bar U}}$, which is entanglement non-increasing as it is a LOCC
operation.

Since ${\cal{T}}^{U\otimes {\bar
U}}(\rho_{\lambda(\nu)})=\sigma(f)$ it follows that
\begin{equation}\label{eq1}
  E_f(\rho_{\lambda(\nu)})\geq E_f(\sigma(f)).
\end{equation}
Now, consider the pure states giving rise to the minimum in the
optimization problem of $\epsilon$  for isotropic states
\cite{TV}:
\begin{equation}\label{eq2}
  \phi_\nu=\sqrt{\nu}\Psi_0+\sqrt{1-\nu} \ket{00}
\end{equation}
By twirling over the discrete group G, i.e., ${\cal
T}(\ketbra{\phi_\nu}{\phi_\nu})=\rho_{\lambda(\nu)}$ we get
$E_f(\rho_{\lambda(\nu)})\leq \epsilon(\sigma(f))$, and thus
\begin{equation}\label{eq3}
\epsilon(\sigma(f)) \geq E_f(\rho_{\lambda(\nu)})\geq
E_f(\sigma(f)) =\text{co}\;\epsilon(\sigma(f)).
\end{equation}
Since the convex hull of $\epsilon$ is the largest convex function
below $\epsilon$ and $E_f(\rho_{\lambda(\nu)})$ is a convex
function, we have indeed $E_f(\rho_{\lambda(\nu)})=
E_f(\sigma(f))$.


\begin{references}
\bibitem{distent} More precisely, it is shown in \cite{VWhash} and \cite{Rains2}
that an entanglement assisted LOCC protocol resp. a PPT preserving
protocol attains the rate $\log d- S(\rho_{\lambda})$, which is in
turn equal to the relative entropy of entanglement. The latter is,
however, a general upper bound for $E_D^+$. Hence, the $E_D^+$ is
in this case equal to the (LOCC/PPT) entanglement assisted
distillable entanglement.
 Moreover, it is
shown in \cite{VWhash}, that for prime dimensions
$E_D^+(\rho_{\lambda})=E_D(\rho_{\lambda})$, which is probably
also true for non-prime dimensions.
\bibitem{PPT} The partial transpose of a density matrix, introduced in the context of
entanglement theory in \cite{PH}, is defined w.r.t. to some
product basis by $\langle ij|\rho^{T_A}|kl\rangle=\langle
kj|\rho|il\rangle$. The set of operations preserving the
positivity of the partial transpose is considerably larger than
the set of LOCC operations.
\end{references}
\end{document}